\documentclass{article}

\usepackage{amsmath}
\usepackage{amssymb}
\usepackage{hyphenat}
\usepackage{epic,eepic}

\usepackage{fullpage}
\usepackage{url}
\usepackage[noend]{algorithmic}

\usepackage{algorithm}
\usepackage{graphicx}

\usepackage{multirow}

\title{Finding the Minimal DFA of Very Large Finite State Automata
  with an Application to Token Passing Networks}

\author{
Vlad~Slavici$^\dag$\and
Daniel~Kunkle$^\dag$\and
Gene~Cooperman$^\dag$\thanks{This
		work was partially supported by
                the National Science Foundation under Grant
                CCF 0916133.}
\and
Stephen~Linton$^\ddag$
\and
\qquad $^\dag\,$Northeastern~University\\
\qquad Boston,~MA\\
\qquad\qquad\qquad\{vslav,kunkle,gene\}@ccs.neu.edu\qquad\qquad\qquad
\and
$^\ddag\,$University~of~St.~Andrews\\
St.~Andrews,~Scotland\\
sal@cs.st-andrews.ac.uk
}

\date{}

\begin{document}
\maketitle

\advance\baselineskip by -0.00pt

\begin{abstract}
Finite state automata (FSA) are ubiquitous in computer science.  Two of
the most important algorithms for FSA processing are the conversion of
a non-deterministic finite automaton (NFA) to a deterministic finite
automaton (DFA), and then the production of the unique minimal DFA for
the original NFA.  We exhibit a parallel disk-based algorithm that
uses a cluster of 29~commodity computers to produce an intermediate
DFA with almost two billion states and then continues by producing
the corresponding unique minimal DFA with less than 800,000 states.
The largest previous such computation in the literature was carried
out on a 512-processor CM-5 supercomputer in 1996.  That computation
produced an intermediate DFA with 525,000 states and an unreported number
of states for the corresponding minimal DFA.  The work is used to provide
strong experimental evidence satisfying a conjecture on a series of token
passing networks.  The conjecture concerns stack sortable permutations
for a finite stack and a 3-buffer.  The origins of this problem lie in
the work on restricted permutations begun by Knuth and Tarjan in the
late 1960s.  The parallel disk-based computation is also compared with
both a single-threaded and multi-threaded RAM-based implementation using
a 16-core 128~GB large shared memory computer.
\end{abstract}

\section{Introduction}
\label{sec:Introduction}

Finite state automata (FSA) are ubiquitous in mathematics and computer
science, and have been studied extensively since the 1950s.  Applications
include pattern matching, signal processing, natural language
processing, speech recognition,
token passing networks (including sorting networks),
compilers, and digital logic.

This work attempts to relieve the critical bottleneck in many automata-based
computations by providing a scalable disk-based parallel algorithm for
computing the minimal DFA accepting the same language as a given NFA. This
requires the construction of an intermediate non-minimal DFA whose, often very
large, size has been the critical limitation on previous RAM-based
computations.  Thus, researchers may use a departmental cluster or a SAN
(storage area network) to produce the desired minimal DFA off-line, and then
embed that resulting small DFA in their production application.  As a
motivating example, Section~\ref{sec:experiment} demonstrates the production of
a two-billion state DFA that is then reduced to a minimal DFA with less than
800,000 states --- a more than 1,000-fold reduction in size.

As a measure of the power of the technique, we demonstrate an application to
the analysis of a series of token passing networks, for which we are now able
to complete the experiments needed to conjecture the general properties of the
whole series and the infinite "limit" network.

This disk-based parallel algorithm is based on a RAM-based parallel algorithm
used on supercomputers of the 1990s.  We adapt that algorithm both to clusters
of modern commodity computers and to a multi-threaded algorithm for modern
many-core computers.  More important, we apply a disk-based parallel computing
approach to carry out large computations whose intermediate data would not
normally fit within the RAM of commodity clusters.  By doing so, we use the
subset construction to produce a 2~billion-state intermediate DFA, and then
reduce that to a minimal DFA of 3-quarters of a million states.  Part of the difficulty of
producing the 2~billion-state DFA by the subset construction is that each DFA
state consists of a subset that includes up to 20 of the NFA states.  Hence, each DFA state needs a
representation of 80~bytes ($4\times 20$).

The novel contributions of this paper are:

\begin{itemize}
\item efficient parallel disk-based versions of known algorithms
for determinizing large NFAs (the subset construction) and for
minimizing very large DFAs;
\item a new multi-threaded implementation for the two algorithms above;
\item an application to challenge problems involving stack-sortable
permutations encoded as token passing networks; and
\item formulation of a conjecture for a series of stack-sortable
permutation problems, based on the experimental evidence arising
from application to that challenge problem.
\end{itemize}

This work represents an important advance over the previous state of
the art~\cite{ravikumar}, which used a 512-processor CM-5
supercomputer to minimize a DFA with 525,000 states.

In the rest of this paper, Section~\ref{sec:RelWork} presents related
work.  Section~\ref{sec:background} presents background on finite state
automata and their minimization.  It also motivates the importance of
the two algorithms (determinization and minimization) by recalling that
NFA and DFA form the primary computationally tractable representations
for the very important class of regular languages in computer science.

Section~\ref{sec:NFADFA} then presents the disk-based parallel
algorithm for determinization (subset construction) and minimization.
It also presents a corresponding multi-threaded computation.
Section~\ref{sec:Application} presents token passing
networks and the challenge problem considered here.
Section~\ref{sec:experiment} presents the experimental results for the
given challenge problem.

\section{Related Work}
\label{sec:RelWork}

Finite state machines are also an important tool in natural language
processing, and have been used for a wide variety of problems in
computational linguistics.  In a work presenting new applications of
finite state automata to natural language processing~\cite{moh96a}, Mohri
cites a number of examples, including: lexical analysis~\cite{sil93};
morphology and phonology~\cite{kos96}; syntax~\cite{moh94, roc96};
text-to-speech synthesis~\cite{spr95}; and speech recognition~\cite{
moh02,per94}.  Speech recognition, in particular, can benefit from the use of
very large automata. In~\cite{moh97}, Mohri predicted: \begin{quote}
{\it ``More precision in acoustic modeling, finer language models,
large lexicon grammars, and a larger vocabulary will lead, in the near
future, to networks of much larger sizes in speech recognition. The
determinization and minimization algorithms might help to limit the size
of these networks while maintaining their time efficiency.''} \end{quote}

While the subset construction for determinization has been a standard
algorithm since the earliest years, this is not true for the minimization
algorithm.  For any DFA there is an equivalent minimal canonical
DFA~\cite[Chapter~4.4]{Hopcroft:2006}.  Fast sequential RAM-based
DFA minimization algorithms have been developed since the 1950s. A taxonomy of
most of these algorithms can be found in~\cite{watson95:taxon}.  The first DFA
minimization algorithms were proposed by Huffman~\cite{Huffman1954161} and
Moore~\cite{Moore1527182}.  Hopcroft's minimization
algorithm~\cite{Hopcroft:1971} is proved to achieve the best possible
theoretical complexity ($O(|\Sigma| N log N)$ for alphabet~$\Sigma$ and
number of states~$N$).  Hopcroft's algorithm has been
extensively revisited~\cite{berstel05, Gries:1972,
Knuutila:2001}. There exist alternative DFA minimization algorithms, such as
Brzozowski's algorithm~\cite{champarnaud}, which, for some
special cases, performs better in practice than Hopcroft's
algorithm~\cite{TabakovVardi}.  However, none of these sequential algorithms
parallelize well (with the possible exception of Brzozowski's, in some cases).

Parallel DFA minimization has been considered since the 1990s. All existing
parallel algorithms are for shared memory machines, either using the CRCW PRAM
model~\cite{tewari}, the CREW pram model~\cite{jaja}, or the EREW PRAM
model~\cite{ravikumar}.  All of these algorithms are applicable for
tightly coupled parallel machines with shared RAM and they make heavy use of
random access to shared memory.  In addition, \cite{ravikumar} minimized
a 525,000-state DFA on the CM-5 supercomputer.

When the DFA considered for minimization is very large (possibly
obtained from a large NFA by subset construction), it must be stored
on disk.  To our knowledge, this work represents the first disk-based
algorithm for determinization and minimization.

Obtaining a minimal canonical DFA equivalent to a given NFA is important for
the analysis of the classes of permutations generated by token passing in
graphs.  Such a graph is called a {\em token passing network}
(TPN)~\cite{Atkinson:1998,Atkinson:1997}. This is related to the subject
permutations~\cite{Atkinson:1999}, with origins in the 1969 work of
Knuth~\cite[Section~2.2.1]{Knuth:1969} and the 1972 work of
Tarjan~\cite{Tarjan:1972}. TPNs are used to model or approximate a range of
data structures, including combinations of stacks, and provide tools for
analyzing the classes of permutations that can be sorted or generated using
them.  Stack sorting problems have been the subject of extensive
research~\cite{Bona:2003}.
 Sorting with
two ordered stacks in series is detailed in~\cite{Atkinson:2002}.
Permutation classes defined by TPNs are described in~\cite{Waton}.
Very recent work focused on permutations generated by stacks and
dequeues~\cite{Albert2010:stacks}.  A collection of results on
permutation problems expressed as token passing networks is
in~\cite{CollectionPermutations}.

\section{Terminology and Background}
\label{sec:background}
\label{sec:FSAPrelim}

Finite state automata and the closely related concepts of regular
languages and regular expressions form a crucial part of the
infrastructure of computer science.  Among the rich variety of
applications of these concepts are natural language grammars, computer
language grammars, hidden Markov models, digital logic, transducers,
models for object-oriented programming, control systems, and speech
recognition.

This section motivates the need for efficient, scalable algorithms
for {\em finite state automata} (FSA), by noting that they are usually
the most computationally tractable form in which to analyze the regular
languages that arise in many branches of computer science.  That analysis
requires efficient algorithms both for determinization of NFA (conversion
of NFA to DFA) and minimization of DFA.

Recall that a {\em deterministic finite state automaton} (DFA) consists
of a finite set of states with labelled, directed edges between pairs
of states.  The labels are drawn from an associated alphabet.  For each
state, there is at most one outgoing edge labelled by a given letter from
the alphabet.  So, a transition from a state dictated by a given letter
is {\em deterministic}.  There is an initial state and also certain of the
states are called {\em accepting}.  The DFA accepts a word if the letters
of the word determine transitions from the initial state to an accepting
state.  The set of words accepted by a DFA is called a {\em language}.

A {\em non-deterministic finite state automaton} (NFA) is similar,
except that there may be more than one outgoing edge with the same
label for a given state.  Hence, the transition dictated by the specified
label is non-deterministic.  The NFA accepts a word if there exists
a choice of transitions from the initial state to some accepting state.

More formally, a DFA is a {\em 5-tuple} $(\Sigma, Q, q_0, \delta, F)$,
where $\Sigma$ is the input alphabet, $Q$ is the set of states of the
automaton, $q_0 \in Q$ is the initial state, and there is a subset of
$Q$, called the {\em final} or {\em accepting} states, $F$. $\delta:
Q \times \Sigma \rightarrow Q $ is the {\em transition function}, which
decides which state the control will move to from the current state upon
consuming a symbol in the input alphabet.

An NFA is a {\em 5-tuple} $(\Sigma, Q, q_0, \delta, F)$. The only
difference from a DFA is that $\delta: Q \times \Sigma \rightarrow
\mathfrak{P} (Q) $.  Upon consuming a symbol from the input alphabet,
an NFA can non-deterministically move control to any one of the defined
next states.

Recall that the {\em subset construction} allows one to transform an
NFA into a corresponding DFA that accepts the same words.  Each state of
the DFA is identified with a subset of the NFA states.  Given a state~A
of the DFA and an edge with label~$\alpha$, the destination state~B
consists of a subset of all states of the NFA having an incoming edge
labelled by $\alpha$ and a source state that is a member of the subset~A.

Finite state automata are an important computationally
tractable representation of {\em regular languages}.  This class of
languages has a range of valuable closure properties, including under
concatenation, union, intersection, complementation, reversal and the
operations of (not necessarily deterministic) transducers.  (A {\em
transducer} is a DFA or NFA that also produces output letters upon
each transition.)  The above properties have algorithmic analogues that
operate on finite state automata. For instance, given an FSA representing
a language it is easy to construct one for the reversed language.  So,
one can compute various operations on regular languages by computing
the analogous operations on their finite state automaton representations.

Using these operations to manipulate regular languages forces one to
choose between a DFA and an NFA representation.  But neither
representation suffices.  Some of the above operations on finite state
automata, such as complementation, require input in the form of a~DFA.
And yet, some operations may transform a DFA into an NFA.

From a computability standpoint, there is no problem.  The subset
construction converts between an NFA and the more specialized DFA.
But while the subset construction is among the best known algorithms of an
undergraduate curriculum, it may also lead to an exponential growth in
the number of states.  This is usually the limiting factor in determining
what computations are practical.

In some cases this problem is completely unavoidable, since there
are families of non-deterministic automata whose languages cannot be
recognized by any deterministic automata without exponentially many
states. In many cases of interest, however, much smaller equivalent
deterministic automata do exist.  But the determinization process alone
is not enough to reduce the DFA to the equivalent unique minimal DFA.
No method is known of finding this minimal DFA without first constructing
and storing the large intermediate DFA.  It is this large intermediate
data which motivates us to consider parallel disk-based computing.

Hopcroft~\cite{Hopcroft:1971} provided an efficient $O(n\log n)$
algorithm for DFA minimization, but the algorithm does not adapt well to
parallel computing.  An efficient parallel $O(n\log^2 n)$ algorithm has
been used in the 1990s~\cite{ravikumar}, but ultimately the lack of
intermediate storage
for the subset construction has prevented researchers from adapting
these techniques for use within the varied applications described above.

\section{Algorithms for NFA to Minimal DFA}
\label{sec:NFADFA}

This section presents disk-based parallel algorithms for both determinization
(Section~\ref{subsec:SubsetConstr}) and DFA minimization
(Section~\ref{subsec:MinimalDFA}), both using {\it streaming} access to data
distributed evenly across the parallel disks of a cluster.  This
avoids the latency penalty that a random access to disk incurs.
Separately, Section~\ref{subsec:MTAlg} presents a depth-first based
algorithm for determinization and minimization suitable for large
shared-memory computers.

For the parallel disk-based implementation, Roomy~\cite{Roomy,RoomyTutorial}
was used.  Roomy is an open-source library for parallel disk-based computing,
providing an API for operations with large data structures. Projects involving
very large data structures have previously been successfully developed using various versions of Roomy:
a parallel disk-based binary decision diagram package~\cite{Kunkle:2010}, or
a parallel disk-based computation, which was used in 2007 to prove
that any configuration of Rubik's cube can be solved in 26 moves
or less~\cite{Kunkle:2007}.
The three Roomy data structures we used are the Roomy hash table, the Roomy list
and the Roomy array.  Each is a distributed data structure, which Roomy keeps
load-balanced on the parallel disks of a cluster. Operations to these data
structures are batched and delayed until the user decides that there are enough
operations for processing to be performed efficiently. In doing so, a latency
penalty is paid only once for accessing a large chunk of data, and aggregate
disk bandwidth is significantly increased.

\subsection{Subset construction for large NFAs}
\label{subsec:SubsetConstr}

For subset construction on parallel disks, three Roomy-hash tables are used:
$visited$, $frontier$ and $next\_frontier$. Hash table keys are sets of states
of the NFA, and hash table values are unique integers.  A hash table entry is
denoted as $(key \rightarrow value)$. A Roomy-list of $triples$ ($set_{id}$,
$transition$, $next\_set_{id}$) is also used, to keep the already discovered
part of the equivalent DFA.

The portion of any Roomy data structure $d$ kept by a specific compute node $k$
is denoted as $d^k$.  Any Roomy data structure is load-balanced across the
disks, so $d^k$ and $d^j$ will be of about the same size, for any compute nodes
$k$, $j$.

Data that needs to be sent to other compute nodes by Roomy is first buffered in
local RAM, in {\it buckets} corresponding to each compute node. For a given
piece of data, Roomy uses a hash function to determine which compute node
should process that data and, hence, in which bucket to buffer that data. Once
a given buffer is full, the data it contains is sent over the network to the
corresponding compute node (or to the local disk, if the data is to be
processed by the local node).

\begin{algorithm}[htb]
\caption{Parallel Disk-based Subset Construction}
\label{alg:ParDiskSubset}
\begin{algorithmic}[1]

  \REQUIRE Initial NFA $init_{NFA}$, with initial state $s_i$ and
  accepting states $F_s$, will be loaded in RAM on each of the
  $N$ compute nodes.
  \ENSURE DFA $interm_{DFA}$, equivalent to $init_{NFA}$

    \STATE Insert $(s_i \rightarrow new \; Id())$ in $visited$ and $frontier$.
    $next\_frontier$ is empty.
    \STATE Each compute node $k$ of the cluster does:
    \WHILE{ $frontier^k$ is not empty }
        \STATE \COMMENT{Compute Neighbors of Frontier}
        \FOR{ each $(set \rightarrow set_{id}) \in frontier^k$ }
            \FOR{ each transition $T$ of $init_{NFA}$ }
            \STATE  Apply $T$ to each NFA state in $set$, to generate $next\_set$.
            \STATE  $next\_set_{id} \leftarrow new \; Id()$
            \STATE Calculate $node$, the compute node responsible for the new
            NFA state, using a hash function $(1 \leq node \leq N)$.
            \STATE Insert $(next\_set \rightarrow next\_set_{id})$ in a local
            RAM-based buffer $sets_{node}$.
            \STATE Insert triple $(set_{id},T,next\_set_{id})$ in a
            local RAM-based buffer $triples_{node}$.
            \ENDFOR
        \ENDFOR
        \STATE \COMMENT{Scatter-Gather, when buffers are full}
        \FOR{ $k \in \{1 \dots N \}$ }
            \STATE Send $sets_{k}$ and $triples_{k}$ to compute node $k$.
        \ENDFOR
        \FOR{ $k \in \{1 \dots N \}$ }
            \STATE Receive a bucket of triples and a buckets of sets
            from each compute node $k$.
        \ENDFOR
        \STATE \COMMENT{Duplicate Detection}
        \STATE Aggregate received $sets$ buffers in $next\_frontier^k$.
        \STATE Remove duplicate and previously visited sets from
        $next\_frontier^k$.
        \STATE  Update all triples that correspond to a duplicate set.
        \STATE Add $next\_frontier^k$ to $visited^k$ and add all $triples$ buffers
        to $triples^k$.
        \STATE $frontier^k \leftarrow next\_frontier^k$
    \ENDWHILE
    \STATE Roomy-list $triples$ now holds $interm_{DFA}$.
    \STATE Convert Roomy-list $triples$ into a compact Roomy-array-based
    DFA representation.
\end{algorithmic}
\end{algorithm}

Parallel disk-based subset construction is described in Algorithm~\ref{alg:ParDiskSubset}.
Parallel breadth-first search (BFS) is used to compute the states of the intermediate DFA.
Duplicate states in each BFS frontier are removed by delayed duplicate detection.
The parallel disk-based computation follows a {\em scatter-gather} pattern in a
loop: local batch computation of neighbors; send results of local computation
to other nodes; receive results from other nodes; and perform duplicate
detection. All parallel disk-based algorithms presented here
(Algorithms~\ref{alg:ParDiskSubset}, \ref{alg:ParDiskMinDFA} and~\ref{alg:ParDiskPartCol})
 use this kind of scatter-gather pattern.

\subsection{Finding the Unique Minimal DFA}
\label{subsec:MinimalDFA}

The algorithm used for computing the minimal DFA on parallel disks is
based on a parallel RAM-based algorithm used on supercomputers in the late
1990s and early 2000s~\cite{jaja,ravikumar,tewari}. We
call this the {\em forward refinement} algorithm.  The central
idea of the algorithm is to iteratively partition the states
(to refine partitions of the states) of the given DFA,
which is proven to converge to a
stable set of partitions.  Upon convergence, the set of partitions,
together with the transitions between partitions, form a graph which is
isomorphic to the minimal DFA. Initially, the DFA states are split into
two partitions: the accepting states and the non-accepting states. A
hash table of visited partitions, $parts$, is used, with pairs of integers as
keys and integers as values.  For the pair of integers, the first
integer represents the partition number of the current state~$i$, while the
second integer represents the partition number of $DFA[i][T]$, where $T$ is
the current transition being processed.  If two states $i$ and~$j$ in the
DFA are equivalent, then for any transition~$T$, at any time during the
iterative process, the pairs corresponding to $i$ and $j$ for the same~$T$
should have the same first integers and the same second integers.
Algorithm~\ref{alg:FwdRef} describes a sequential RAM-based version
of {\em forward refinement}, while Algorithm~\ref{alg:ParDiskMinDFA}
describes the parallel disk-based one.

\begin{algorithm}[hbt]
\caption{Sequential RAM-based Forward Refinement}
\label{alg:FwdRef}
\begin{algorithmic}[1]

  \REQUIRE A DFA $init_{DFA}$, with $N$ states,
  with initial states $I_s$ and accepting states $F_s$
  \ENSURE The minimal canonical DFA $min_{DFA}$, equivalent to $init_{DFA}$.

  \STATE Initialize array $curr\_refs$: $curr\_refs[i] \leftarrow 0$
  if $i$ is a non-accepting state of $init_{DFA}$, and
  $curr\_refs[i] \leftarrow 1$ if $i$ is an accepting state.
  \STATE Initialize array $next\_refs$ to all $0$.
  \STATE $prev\_num\_refs \leftarrow 0; curr\_num\_refs \leftarrow 2$
  \WHILE{ $prev\_num\_refs < curr\_num\_refs$ }
    \STATE $prev\_num\_refs \leftarrow curr\_num\_refs$
    \FOR{ each transition T of $init_{DFA}$ }
      \STATE Initialize hash table $parts$ to $\emptyset$
      \STATE $next\_id \leftarrow 0$
      \FOR{ $i \in \{1 \dots N \}$ }
        \STATE $next\_part \leftarrow curr\_refs[init_{DFA}[i][T]]$
	\STATE $pair \leftarrow  new \; Pair(curr\_refs[i], next\_part)$
        \STATE $id \leftarrow parts.getVal(pair)$
        \IF{ $id$ was not found in $parts$ }
          \STATE Insert $(pair \rightarrow$ $next\_id)$ in $parts$
          \STATE $id \leftarrow next\_id$
          \STATE $next\_id \leftarrow next\_id + 1$
        \ENDIF
        \STATE $next\_refs[i] \leftarrow id$
      \ENDFOR
      \STATE $curr\_refs \leftarrow next\_refs$
    \ENDFOR
    \STATE $curr\_num\_refs  \leftarrow   next\_id$
  \ENDWHILE
  \COMMENT{ For each state $i$ of $init_{DFA}$, $curr\_refs[i]$ defines
  what partition state $i$ is in.}
  \STATE Collapse each partition to just one state to
  obtain the minimal DFA.
\end{algorithmic}
\end{algorithm}

The major differences between Algorithms~\ref{alg:FwdRef}
and~\ref{alg:ParDiskMinDFA} are that lines 7--17 and line~20 of
Algorithm~\ref{alg:FwdRef} are parallelized and that Roomy's principles
of parallel
disk-based computing are used: all large data structures are split into
equally-sized chunks which are kept on the parallel disks of a cluster and all
access and update operations to the $curr\_refs$ and $prev\_refs$ arrays and to
the $parts$ hash table are delayed and batched for efficient streaming
access to disk.  Also, duplicate detection,
which in the sequential RAM-based algorithm appears in lines 12--17, is
replaced by delayed duplicate detection.

\begin{algorithm}[hbt]
\caption{Parallel Disk-based Forward Refinement}

\label{alg:ParDiskMinDFA}
\begin{algorithmic}[1]

  \REQUIRE A DFA $init_{DFA}$, with $N$ states,
  with initial states~$I_s$ and accepting states~$F_s$
  \ENSURE The minimal canonical DFA $min_{DFA}$, equivalent to $init_{DFA}$.

  \STATE \COMMENT{Initialization and outer loop are the
   same as lines 1--6 in Algorithm~\ref{alg:FwdRef}}
  \STATE \COMMENT{Disk-based parallel loop (parallelization of lines 7--17
   in Algorithm~\ref{alg:FwdRef}) -- each node $k$ does:}
  \STATE Initialize hash table $parts^k$ to $\emptyset$
    \FOR{ $i \in \{states^k\}$ }
      \STATE $next\_part[i] \leftarrow curr\_refs[init_{DFA}[i][T]]$
      \STATE $pair[i] \leftarrow  new \; Pair(curr\_refs[i], next\_part[i])$
      \STATE $next\_id[i] \leftarrow new \; Id()$
    \ENDFOR
    \FOR{ $i \in \{states^k\}$ }
      \STATE Send new entry $(pair[i] \rightarrow next\_id[i])$ and
      state id $i$ to node $= hash(pair[i])$
    \ENDFOR
    \FOR{ $k \in \{1 \dots N\}$ }
      \STATE Receive $pair \rightarrow id$ entries from node $k$
    \ENDFOR
    \FOR{ each received entry $pair \rightarrow recv\_id$ and associated state id $i$
    from a node $k$ }
      \IF{ an entry $pair \rightarrow id$ was not found in the local $parts$ }
        \STATE Insert $(pair \rightarrow recv\_id)$ in the local $parts$
        \STATE Send key--value pair $i \rightarrow recv\_id$ to node $k$
      \ELSE
        \STATE Send key--value pair $i \rightarrow id$ to node $k$
      \ENDIF
    \ENDFOR
    \FOR{ $k \in \{1 \dots N\}$ }
      \STATE Receive $i \rightarrow id$ entries from node $k$
    \ENDFOR
    \FOR{ each received entry $i \rightarrow id$ }
      \STATE $curr\_refs[i] \leftarrow id$
    \ENDFOR

\end{algorithmic}
\end{algorithm}

Note that in Algorithm~\ref{alg:ParDiskMinDFA} each compute node $k$ keeps
its own part of the $parts$ hash table ($parts^k$) and owns a part of the
intermediate DFA states ($states^k$). As with subset construction,
the parallel disk-based computation follows a scatter-gather pattern,
denoted in the pseudocode by most of the {\em for} loops:
local computation (lines 4--7), {\em scatter} (lines 8--9),
{\em gather} (lines 10--11),
local computation and {\em scatter} (lines 12--17),
{\em gather} (lines 18--19) and local computation (lines 20--21).

The last part of finding the minimal DFA, in which each partition
collapses to one state, is presented separately,
in Algorithm~\ref{alg:ParDiskPartCol}.

\begin{algorithm}[hbt]
\caption{Parallel Disk-based Partitions Collapse}

\label{alg:ParDiskPartCol}
\begin{algorithmic}[1]

  \STATE \COMMENT{Collapsing partitions to $min_{DFA}$ (parallelization
   of line~20 in Algorithm~\ref{alg:FwdRef}) -- each node $k$ does:}
  \FOR{ $i \in \{indices^k\}$ }
    \STATE Get $partition[i]$ (the partition of state $i$)
    from $curr\_refs^k$
    \FOR{ each transition T of $init_{DFA}$ }
      \STATE Get $partition[init_{DFA}[i][T]]$ from node that owns it
    \ENDFOR
    \STATE \COMMENT{Now all the transitions of $partition[i]$ in
     $min_{DFA}$ are known}
  \ENDFOR

\end{algorithmic}
\end{algorithm}

\subsection{Multi-threaded Implementations for Shared Memory}
\label{subsec:MTAlg}

For comparison with the parallel disk-based algorithms, multi-threaded
shared-memory implementations of subset construction and DFA
minimization are provided.  A shared-memory architecture almost always
has less storage (128~GB RAM in our experiments) than parallel disks.
To alleviate the state combinatorial explosion issue, depth-first
search (DFS) is used here for the subset construction instead of
breadth-first search~(BFS).

The smallest instance of the four NFA to minimal DFA problems considered can be
solved on a commodity computer with 16~GB of RAM, the second instance needs 40
GB of memory for subset construction, while the third largest instance needs a
large shared-memory machine with at least 100~GB of RAM. The largest instance
considered cannot be solved even on a large shared-memory machine, thus
requiring the use of parallel disks on a cluster.

A significant problem for both subset construction and DFA minimization
in a multi-threaded environment is synchronization for duplicate detection.
For subset construction, this issue arises when a thread discovers a new DFA
state and checks whether the state has been discovered before by itself or
another thread. The data structure keeping the already-discovered states
(usually a hash table) has to be
locked in that case, so that the thread can check whether the current state
has already been discovered and, if not, to insert the new state in the
data structure. However, such an approach would lead to excessive
lock contention (many threads waiting on the same lock).

Hence, the solution employed was to use a partitioned hash table to keep the
already discovered states instead of a regular hash table.
For large problem instances, the hash table was partitioned into
1024~separate hash tables --- each with its own lock.
So long as the number of partitions
is much larger than the number of threads, it is unlikely that two
threads will concurrently discover two states that will belong to the
same partition, thus avoiding most of the lock contention.
Experiments (see Section~\ref{sec:experiment},
Table~\ref{tab:MTStackDepth11})
show significant speedup with the increase in number of threads.

A similar solution was used for the forward refinement algorithm, which
minimizes the DFA obtained from subset construction.  In this case,
read accesses significantly dominate over write accesses.
The implementation took advantage of this by implementing a lock
only around writes to the corresponding hash table.  The valid
bit was written last in this case.  A write barrier is needed to guarantee
no re-ordering of writes.  In the worst case, a concurrent read may
read the hash entry as invalid, and that thread will then request
the lock, verify that the hash entry is still invalid, and if that
is the case, then do the write.  This is safe.

\section{Token Passing Networks}
\label{sec:Application}

Section~\ref{sec:TPN} provides background on token passing networks,
and the specific challenge problem addressed here.
Section~\ref{sec:TPNcomputation} describes the computation on token
passing networks addressed here and the component of that computation
that requires the parallel solutions of this paper.

\subsection{Stacks, Token Passing Networks and Forbidden Patterns}
\label{sec:TPN}

The study of what permutations of a stream of input objects could be achieved
by passing them through various data structures goes back at least to
Knuth~\cite[Section~2.2.1]{Knuth:1969}, who considered the case of a stack
and obtained a simple
characterization and enumeration in this case.  Knuth's characterization uses
the notion of \textit{forbidden substructures}: a permutation can be achieved
by a stack if and only if it does not contain any three numbers (not
necessarily consecutive) whose order within the permutation, and relative
values match the pattern  high-low-middle (usually written~312). For instance
41532 cannot be achieved because of~4, 1 and~2.  This work has spawned a
significant research area in combinatorics, the study of
permutation patterns~\cite{CollectionPermutations} in which much beautiful
structure has been revealed. Nevertheless, many problems very close to Knuth's
original one remain unresolved: in particular there is no similar
characterization or enumeration of the permutations achievable by two stacks in
series (it is not even known if 2-stack achievability can be tested in
polynomial time).  A number of authors have investigated restricted forms of
two-stack achievability~\cite{Bona:2003} including the case of interest here,
where the stacks are restricted to finite capacity, in which case they can be
modelled as \textit{token passing networks},
as introduced in~\cite{Atkinson:1997}.

To recap briefly, a token passing network is a directed graph with
designated input and output vertices.  Numbered tokens are considered
to enter the graph one at a time at the input vertex, and travel along
edges in the appropriate direction. At most one token is permitted at
any vertex at any time. The tokens leave the graph one at a time at the
output vertex. A permutation $\pi \in S_n$ is called \textit{achievable}
for a given network if it is possible for tokens to enter in the order
$1,\ldots,n$ and leave in the order $1\pi,\ldots,n\pi$.

\begin{figure}[hbt]

\begin{center}
\includegraphics{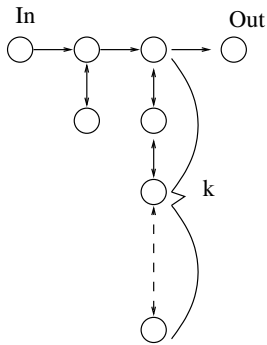}
\end{center}
\caption{A 2-stack followed by a $k$-stack represented as a token passing network}
\label{fig:twokstack}
\end{figure}

 In this case, the two
stacks can be modelled as a finite token passing network (as seen,
for example in Figure~\ref{fig:twokstack}) and their behavior studied
using the techniques of~\cite{Atkinson:1997}. These techniques allow
the  classes of achievable permutations and the forbidden patterns that
describe them to be encoded by regular languages and manipulated using
finite state automata using a collection of GAP~\cite{GAP4} programs
developed by the fourth author and M. Albert.

In previous work, the fourth author explored the cases of stacks of depths~2
and depth~$k$ (as seen in Figure~\ref{fig:twokstack}) for a range of values
of~$k$ and observed that for large
enough~$k$ the sets of minimal  forbidden patterns appeared to converge to a
set of just~20 of lengths between 5 and~9, which were later
proved~\cite{Elder:2006} to describe the case of a 2-stack and
an infinite stack.

\begin{figure}[htb]

\begin{center}
\includegraphics{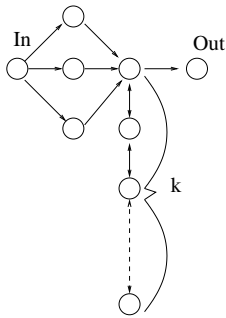}
\end{center}
\caption{A 3-buffer followed by a $k$-stack represented as a token passing network}
\label{fig:bufandstack}\end{figure}

The application that motivates the calculations in this paper is a step
towards extending this result to a 3-stack and an infinite stack, by way of
the slightly simpler case of a 3-buffer (a data structure which can hold up to
three items and output any of them). This configuration is shown in
Figure~\ref{fig:bufandstack}.

 Computations had been completed on
various sequential computers for a 3-buffer and a $k$-stack for $k \le 8$, but
this was not sufficient to observe convergence. The examples considered in this
paper are critical steps in the computations for
$k = 9$, $k=10$, $k=11$ and $k=12$.
Based on the results of these computations we are now able to conjecture with
some confidence a minimal set of 12,636 forbidden permutations for a 3-buffer
and an infinite stack of lengths between 7 and~18.

\subsection{The Computation}
\label{sec:TPNcomputation}

The computations required for these investigations are those implied by
Corollary~1 of~\cite[p.~96]{Albert:2001}. By modelling the token passing
network in the style of~\cite{Atkinson:1997}, slightly optimized to avoid
constructing so many redundant states, we can construct (an automaton
representing) a language $L$ describing the permutations achievable by our
network, and we wish to construct a language $B$ describing the minimal
forbidden patterns.  Each state of $L$ represents a configuration of the
network and the labels on the transitions represent (rank encoded) output
symbols, if any. Combining results from~\cite{Albert:2001} and simplifying the
notation a little we find

$$B = \left( L^{RC} \cap \left( L^{RC} D^t\right)^C\right)^R$$

\noindent
where $R$ denotes left-to-right reversal, $C$ denotes complementation and $D$
is the deletion transducer described in~\cite{Albert:2001}. Each step of this
computation can be realized by standard algorithms using finite state automata,
but, as observed above, with frequent recourse to determinization (to allow
complements) and minimization (to control explosion in the number of states).
As the computations become larger, the limiting step turns out to arrive after
the application of the transposed  deletion transducer and before the next
complementation, and it is this step that we have parallelized in this paper.

\section{Experimental Results}
\label{sec:experiment}

\subsection{Parallel Disk-based Computations}
\label{subsec:ExpParDisk}

Parallel disk-based computations were carried out on a 29-node computer
cluster, each node's processor being a 4-core Intel Xeon CPU 5130 running at
2 GHz. Nodes on the cluster had either 8 or 16~GB of RAM and at least 200~GB
of free disk storage and ran Red Hat Linux kernel version~2.6.9.

Table~\ref{tab:Results} presents the sizes of the
intermediate DFAs produced by subset construction and the sizes of the minimal
DFAs produced by the minimization process for the four considered
token passing network problems (corresponding to stack depths 9, 10, 11 and~12).

\newcommand\T{\rule{0pt}{3.1ex}}

\newcommand\B{\rule[-1.7ex]{0pt}{0pt}}

\begin{table}[htb]\small
\centering
\caption{Solutions for the four considered problems.}
  \addtolength{\tabcolsep}{-3pt}
  \begin{tabular}{|r|r|r|r|}
   \hline 
      \textbf{Stack} \T & \textbf{NFA size}  & \textbf{Interm. DFA} &\textbf{Min. DFA}  \\
      \textbf{depth}    & (\#states) & \textbf{size} (\#states) & \textbf{size} (\#states)\\
  \hline\hline
        9\T             & 167,143       & 49,722,541            & 32,561 \\
       10               & 537,294       & 175,215,168           & 95,647 \\
       11               & 1,667,428     & 587,547,014           & 274,752\\
       12               & 5,035,742     & 1,899,715,733         & 774,172\\

   \hline
  \end{tabular}
\label{tab:Results}
\end{table}

Table~\ref{tab:ParDiskSubset} shows the running time and aggregate
disk-space used by the subset construction results for the four
problem instances.  Each state in the intermediate DFA is a subset
of states in the original NFA and needs to be kept as such until the
subset construction phase is over, for the purpose of exact duplicate
detection. Hence, for each newly discovered DFA state, the entire
corresponding subset needs to be stored on disk. The average subset size
(the sum of all subset sizes divided by the number of subsets) increases
slightly with stack depth, from an average of 8.48~states per set for
stack depth~8 to 10.06~states per set for stack depth~12.

\begin{table}[htb]
\centering
\caption{Parallel disk-based subset construction.}
  \small{
  \addtolength{\tabcolsep}{-3pt}
  \begin{tabular}{|r|r|r|r|c|}
   \hline 
      \textbf{Stack} \T & \textbf{NFA size}  & \multicolumn{3}{|c|}{\textbf{Intermediate DFA}} \\ \cline{3-5}
      \textbf{depth} & (\#states) & \textbf{Size} (\#states) & \textbf{Peak disk} & \textbf{Time} \\
  \hline\hline
        9\T          & 167,143       & 49,722,541      &  24 GB      & 9min                    \\
       10            & 537,294       & 175,215,168     &  90 GB      & 29min                   \\
       11            & 1,667,428     & 587,547,014     &  327 GB     & 3h 40min                \\
       12            & 5,035,742     & 1,899,715,733   &  1,136 GB   & 1day 12h      \\

   \hline
  \end{tabular}
  }
\label{tab:ParDiskSubset}
\end{table}

\begin{figure}
\begin{center}
\includegraphics[width=0.4\textwidth]{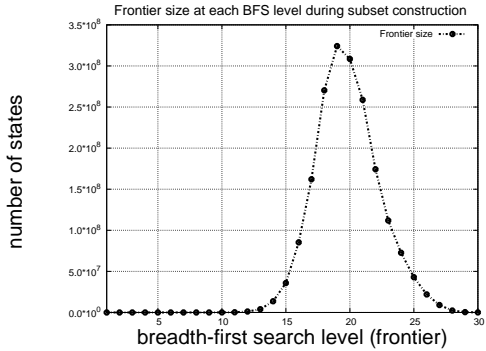}
\end{center}
\caption{Frontier sizes for each BFS level of the implicit graph
corresponding to subset construction.}
\label{fig:bfs}
\end{figure}

Figure~\ref{fig:bfs} presents the breadth-first search frontier sizes
for the largest case ($k=12$).  This and the other three cases exhibit
a thin bell-shaped curve, in contrast to the pear-shaped curve seen
for many other implicit graph enumerations.

The intermediate DFA (produced by subset construction) was then
minimized using the forward refinement algorithm.  Experimental results
for DFA minimization are presented in Table~\ref{tab:ParDiskMinDFA}.
For each of the four problem instances, the computation required exactly
five forward refinements (five of the outer iterations described in
Algorithm~\ref{alg:ParDiskMinDFA}).

\begin{table}[htb]
\centering
\caption{Parallel disk-based DFA minimization results.}
  \small{
  \addtolength{\tabcolsep}{-3pt}
  \begin{tabular}{|r|r|r|r|c|}
   \hline 
      \textbf{Stack} \T & \textbf{Num.}   & \textbf{Interm. DFA} & \multicolumn{2}{|c|}{\textbf{Minimal DFA}} \\ \cline{3-5}
      \textbf{depth}    & \textbf{trans.} & \hfill\textbf{Size} (\#states)\hfill  & \textbf{Peak disk} &  \textbf{Time} \\ \hline
\hline
        9\T             & 11              & 49,722,541           & 6 GB        & 38min            \\
       10               & 12              & 175,215,168          & 22 GB       & 2h 42min               \\
       11               & 13              & 587,547,014          & 81 GB       & 9h 20min                \\
       12               & 14              & 1,899,715,733        & 295 GB      & 1day 8h       \\

   \hline
  \end{tabular}
  }
\label{tab:ParDiskMinDFA}
\end{table}

The DFA minimization times, reported in Table~\ref{tab:ParDiskMinDFA},
grow steadily, almost linearly, with the increase in number of states of
the intermediate DFA.  On the other hand, the subset construction times
from  Table~\ref{tab:ParDiskSubset} increase much more rapidly. There
are two reasons for this.  First, the two smaller cases run faster because the
distributed subset construction fits in the aggregate RAM of the nodes
of the cluster.  Second, we suspect the computation to be network-limited.
The cluster is five years old and uses the 100~Mb/s (12.5~MB/s) Fast
Ethernet commodity network of that time.  This point-to-point
network speed is significantly slower than disk.
This especially penalizes the two larger cases.

\subsection{Multi-threaded RAM-based Computations}
\label{subsec:ExpMtRAM}

Multi-threaded computations were run on a large shared-memory machine with
four quad-core 1.8~GHz AMD Opteron processors (16~cores), 128~GB of RAM,
running Ubuntu~9.10 with a SMP Linux~2.6.31 server kernel.

Only the first three computations could be completed on the large shared-memory
machine used. The fourth computation requires far too much memory.
Table~\ref{tab:MTStackDepth11} shows how the running time
of the subset construction and DFA minimization
scales with the number of worker threads. The reported timings are
for the stack depth~11 problem instance. The size of the intermediate DFA
produced by subset construction for this instance is 587,547,014 states. The
minimal DFA produced by forward refinement has 274,752 states. For
any number of worker threads, the peak memory usage for subset construction
was 98 GB, while for minimization it was 36.5 GB.

\begin{table}[htb]
\centering
\caption{Multi-threaded RAM-based timings for stack depth~11.}
  \small{
  \addtolength{\tabcolsep}{-4.5pt}
  \begin{tabular}{l|c|c|c|c|c|}

      \multicolumn{1}{c}{}  &   \multicolumn{5}{c}{\textbf{Num. threads}}   \\\cline{2-6}
      \textbf{Subset}       &   1 &    2     &    4       &   8       & 16 \\\cline{2-6}
      \textbf{constr.}  \B & 15h 30min  \T & 8h 10min & 3h 50 min  & 2h 5min   & 1h 15min   \\\hline
      \textbf{Minimiz.} \T & 8h  & 5h 5min  & 2h 40 min  & 1h 25min  & 57min   \\\hline\hline
      \textbf{Total} \T & 23h 30min & 13h 15min & 6h 30 min & 3h 30min & 2h 12min   \\\cline{2-6}
  \end{tabular}
  }
\label{tab:MTStackDepth11}
\end{table}

The timings in Table~\ref{tab:MTStackDepth11} show that both
the multi-threaded subset construction and the DFA minimization
implementations scale almost linearly with the number of threads. DFA
minimization scales almost linearly for up to 8 threads. From 8 to 16
threads it scales sub-linearly due to significant lock contention.

Table~\ref{tab:MTStackDepth9-10} presents the timings for the two smallest
instances when using 16 worker threads. For the stack depth~9 case, the peak memory usage was
12~GB for subset construction and 5~GB for DFA minimization.
For stack depth~10, the peak memory usage was 40~GB and 11~GB, respectively.

\begin{table}[htb]
\centering
\caption{Multi-threaded RAM-based results for stack depths 9 \& 10, with 16 worker threads.}
  \small{
  \addtolength{\tabcolsep}{-3pt}
  \begin{tabular}{|r|c|c||c|}
   \hline 
     \textbf{Stack depth} & \multicolumn{3}{|c|}{\textbf{Time}}     \\\cline{2-4}
                          & \textbf{Subset constr.} & \textbf{DFA min.} & \textbf{Total}   \\\hline
      9                   & 8 min          &  4min 10s & 12min 10s \\
     10                   & 25 min         &  15min    & 40min \\
     11                   & 1 hr 15 min    &  57min    & 2hr 12min \\ \hline
  \end{tabular}
  }
\label{tab:MTStackDepth9-10}
\end{table}

\bibliographystyle{abbrv}
\bibliography{slavicietal2011}

\end{document}